\documentclass[12pt]{iopart}
\usepackage{graphicx}

\newcommand{\xv}{$\xi_{\rm v}$}
\newcommand{\msr}{$\mu$SR}
\newcommand{\la}{$\lambda$}

\begin{document}
% Journal identifier can be put here if required, e.g.
%\jl{14}

\title{Field-induced quasiparticle excitations in novel type II superconductors}

\author{R Kadono\dag\ddag}

\address{\dag\ Institute for Materials Structure Science,
High Energy Accelerator Research Organization (KEK),
Tsukuba, Ibaraki 305-0801, Japan}

\address{\ddag\ School of Mathematical and Physical Science,
The Graduate University for Advanced Studies,
Tsukuba, Ibaraki 305-0801, Japan}

\begin{abstract}
We show that the field dependence of the magnetic penetration depth ($\lambda$), 
for which muon spin rotation ($\mu$SR) is an excellent microscopic
probe, provides useful information
on the degree of anisotropy for the superconducting order parameter.
In type II superconductors associated with anisotropic order parameters,
$\lambda$ is sensitive to the quasiparticle excitation 
induced by the Doppler shift due to a supercurrent around magnetic vortices.
The presence of such low-energy excitations manifests itself in the non-zero
slope of $\lambda$ against an external magnetic field.
We review recent results on the field dependence of
$\lambda$ obtained from the application of $\mu$SR 
to novel superconductors that exhibit unconventional
characters associated with the anisotropic order parameter.
\end{abstract}


%\submitted

\maketitle

\section{Introduction}

One of the greatest impacts from the discovery of high-$T_c$
cuprate superconductors is renewed interest in the exotic mechanisms
of superconductivity, and a boosted search for their model systems, which has led
to a soaring number of novel materials identified as superconductors\cite{M2S:00}.
They include varieties of transition-metal
oxides, borides, borocarbides, and other intermetallic compounds
with rare-earth elements.  The list can be readily extended by including
those based on organic materials, and is still growing in length.
Compared with classical
simple metals or binary compound superconductors, they have
a common distinctive feature that the pairing correlation is
potentially highly anisotropic due to a strong electronic correlation (Coulomb
repulsion) and/or a considerably two-dimensional nature of the Fermi surface,
which is also shared by cuprate superconductors.

It is well established that any metallic system can fall into a 
superconducting state when there is an effective interaction that is attractive
between the conduction electrons (pair correlation).  In this situation,
electrons tend to form a bound (pairing) state between those with opposite momenta
(${\bf k}$ and $-{\bf k}$), so that the phase of the pairing
wave-function may become absolute zero\cite{Cooper:56}.
This instability towards the formation
of electronic bound states without a barrier is an intrinsic character of the Fermi
surface, irrespective of the microscopic mechanism of the attractive interaction.
 When the pair correlation is
mediated by the electron-phonon interaction, as in ordinary cubic metals,
the pair correlation has the least dependence on ${\bf k}$
(relative orbital angular momentum $L=0$), and thereby
the structure of the superconducting order parameter $\Delta({\bf k})$ is isotropic
over the entire Fermi surface.  Because of the mandatory requirement
of Fermi statistics that the electronic wave-function must be anti-symmetric,
the remaining freedom of the spin state in the paring function is set to be a singlet ($S=0$).
This pair correlation, having $s$-wave and spin singlet symmetry, is conventionally
called the ``BCS mechanism"; in its narrower sense,
the electron-phonon interaction is presumed to be the primary origin of the attractive
 interaction\cite{Bardeen:57}.

Historically, the first sign of non-BCS type pairing was found in the superfluidity of
liquid $^3$He\cite{Osheroff:72,Legget:73},
where the neutral $^3$He atoms (which have a nuclear spin of 1/2 and
thereby obey Fermi statistics) play the role of electrons in superconductivity.
It is known that liquid $^3$He can be regarded as being a Fermi liquid (i.e., having a
well-defined Fermi surface) below $\sim10^2$ mK, and that it becomes a superfluid
below a few mK, where it can flow narrow channels without friction (superfluidity).
However, there is a major difference between the
nature of pairing between $^3$He atoms and that of the BCS type.
The $^3$He atoms have a hard core of a repulsive interaction with
a relatively large radius, which makes it difficult to pair in a state with
zero angular momentum.  Thus, many theories have predicted that $^3$He atoms
may pair in a $p$-wave ($L=1$) or $d$-wave ($L=2$) state, where they can
keep themselves apart while the pairing interaction is at 
work\cite{Anderson:61,Balian:63}.
Later, experiments confirmed that they were indeed in a $p$-wave state.
This also meant that the spin part of the pairing wave-function
(order parameter) must be
in a triplet state ($S=1$), leading to a variety of possibilities for the total
state of the pairing wave-function ($\propto\Delta({\bf k})$) to break
symmetry.  It is now established that there are at least
three different phases in the superfluidity of liquid $^3$He that all have
different order parameters.

It is readily predicted that the situation similar to that in liquid $^3$He
can be realized when a short-range repulsive electronic
correlation is not negligible in metallic superconductivity.
As has been established
during the past decade, high-$T_c$ cuprates are among the first such examples
in which electrons pair in a state other than an $s$-wave due to a strong
on-site repulsive correlation.  The Cooper pairs
in cuprates prefer a $d$-wave because of the tetragonal structure
of the two-dimensional CuO$_2$ lattice and the associated symmetry of
the Fermi surface.  More specifically, the pairing is a $d_{x^2-y^2}$-wave
and the order parameter is described by a gap function,
\begin{equation}
\Delta({\bf k}) = \Delta_0(\cos k_x-\cos k_y)\simeq \Delta_0(k_y^2-k_x^2),
\end{equation}
where $\Delta_0$ is the maximum value of the anisotropic gap;
the energy gap vanishes along the lines $k_x=\pm k_y$, which are called
{\sl line nodes}.
It also happens that the
electronic correlation in cuprates is antiferromagnetic, as is naturally expected
for doped Mott insulators,  which makes it favorable to form spin singlet
pairs.  The latter points to a magnetic origin as a pairing mechanism, 
irrespective of the true nature of the ground state which is still under debate.
Thus, the structure of the superconducting order parameter reflects important
characteristics of the electronic correlation.

In this paper, we review our recent studies on the structure of the 
superconducting order parameter in novel type II superconductors by
muon spin rotation/relaxation ($\mu$SR).
It is well known that a magnetic field can penetrate type II superconductors
as a bundle of quantum flux lines (magnetic vortices)\cite{Abrikosov:57},
where the spatial field distribution,
$B({\bf r})$, becomes inhomogeneous due to gradual
change in the supercurrent flow around the vortices.  The degree of inhomogeneity, which
is primarily determined by the  magnetic penetration depth ($\lambda$),
a magnetic cutoff parameter ($\xi_{\rm v}$, which is proportional to the Ginsburg-Landau
(GL) coherence length $\xi_{\rm GL}$) and
the spacing of vortices ($a$) can be measured directly by $\mu$SR
as a spin-spin relaxation ($1/T_2$).  By applying a refined analysis technique,
one can reconstruct $B({\bf r})$ more accurately so that both $\lambda$ and $\xi_{\rm v}$
may be deduced separately\cite{Sonier:00}.
Among various experimental techniques applied to a similar end,
$\mu$SR technique is unique in many respects.
Namely, it is a microscopic technique that can be applied to
virtually any superconductors having
a reasonable magnetic penetration depth ($\lambda\le 5000$ \AA).
Because of the purely magnetic nature of a spin 1/2 probe,
the interpretation of the $\mu$SR spectra is free from any complication
due to additional interactions from higher multipoles often found in
nuclear magnetic resonance (NMR).
The $\mu$SR technique is sensitive to the bulk property, and is thus free from
effects specific to the surface, while they are often problems for scanning
tunneling spectroscopy (STS) or angular-resolved photo-emission spectroscopy
(ARPES).

In the following, we demonstrate that the temperature/field dependence of $\lambda$
is strongly affected by the anisotropy of the order parameter.  In particular,
$\lambda$ is enhanced by an external field due to the Doppler shift
of quasiparticles in the gap nodes\cite{Volovik:93},
which leads to almost a linear increase of $\lambda$ with increasing field.
This feature is regarded as an unambiguous sign of the presence of nodes in the
energy gap.
We show several examples of field-dependent $\lambda$, some of which are
indeed identified to have anisotropic order parameter by other experimental
techniques.  A more comprehensive review of a similar study
can be found elsewhere\cite{Sonier:00}.

\section{Internal Magnetic Field Distribution in the Mixed State}

\subsection{Microscopic Model}
In penetration-depth measurements, it is common to assume a geometrical
condition that muons are implanted into a specimen
with the initial spin polarization perpendicular to the external field,
${\bf H}=(0,0,H)$.
Then, since muons stop randomly along the length scale of the flux line lattice (FLL),
the time evolution of complex muon polarization, $\hat{P}(t)$, provides
a random sampling of the internal field distribution, ${\bf B}({\bf r})=(0,0,B({\bf r}))$:
\begin{eqnarray}
\hat{P}(t) &\equiv & P_x(t)+iP_y(t)=\exp(-\sigma_{\rm b}^2t^2)\int_{-\infty}^\infty
n(B)\exp(i\gamma_\mu Bt+\phi)dB,\label{Pt}\\
n(B) & = & \langle\delta(B-B({\bf r}))\rangle_{\bf r},
\end{eqnarray}
where $P_{x,y}(t)$  is proportional to the time-dependent $\mu^+$-$e^+$ decay
asymmetry, $A_{x,y}(t)$, deduced from a corresponding
sets of positron counters,
$n(B)$ is the spectral density for the internal field defined as a spatial
average ($\langle\:\rangle_{\bf r}$) of the delta function,
$\gamma_\mu$ is the muon gyromagnetic ratio (=$2\pi\times$135.53 MHz/T),
and $\phi$ is the initial phase of muon precession\cite{Brandt:88a,Brandt:88b}.
The additional relaxation ($\sigma_{\rm b}$) is mainly due to
random local fields from nuclear magnetic moments
($\sigma_{\rm n}\sim$0.1 $\mu$s$^{-1}$) and
the distortion of flux line lattice due to a random pinning of vortices
($\sigma_p$), which can be approximated by a Gaussian relaxation\cite{Brandt:88a}.
These equations indicate that the real amplitude of
the Fourier-transformed muon
precession signal corresponds to $n(B)$ with an appropriate correction of
$\sigma_{\rm b}$.  While $\sigma_n$ can be estimated from the spectrum
in the normal state, $\sigma_p$ often needs to be considered as a variable
parameter, depending on the temperature and/or external magnetic field.

In modeling the internal field distribution of vortex state,
the simplest approach is to assume that the field distribution is a linear
superposition of that for an isolated vortex, as presumed in the London
theory\cite{London:35}.  This is a reasonable assumption as long as the
inter-vortex distance is much longer than the GL coherence length
($H\ll H_{c2}$). Then, what we need is to know the
field distribution around a single vortex and the structure of the vortex
lattice.  The latter must be given from other sources of information,
such as small angle neutron scattering (SANS) or scanning tunneling
microscopy/spectroscopy (STM/STS).
In the London model, $B({\bf r})$ is approximated as the sum of
the magnetic induction from isolated vortices, to yield
\begin{equation}
B({\bf r})=B_0\sum_{\bf K}\frac{e^{-i{\bf K}\cdot{\bf r}}}
{1+K^2\lambda^2}F(K,\xi_{\rm v})\:,
\label{Br}
\end{equation}
where ${\bf K}$ are the vortex reciprocal lattice vectors,
$B_0$ ($\simeq H$) is the average internal field, $\lambda$ is the
London penetration depth, and $F(K,\xi_{\rm v})$ is a nonlocal
correction term with \xv\ being the cutoff parameter for the {\sl magnetic} field
distribution; care must be taken not to interpret \xv\ naively as $\xi_{\rm GL}$
which is for the spatial variation of the superconducting order parameter.
While the Lorentzian
cutoff, $F(K,\xi_{\rm v})=\exp(-\sqrt{2}K\xi_{\rm v})$, is predicted to be
a better approximation for the GL theory at lower fields\cite{Yaouanc:97},
the Gaussian cutoff, $F(K,\xi_{\rm v})=\exp(-K^2\xi_{\rm v}^2/2)$, generally provides
satisfactory agreement with the data\cite{Brandt:72}.
Note, however, that the Gaussian cutoff is derived
from the GL equations near $H_{c2}$, and thus would not be appropriate at lower fields.
A comparison of the analysis on identical data indicates that a Gaussian cutoff yields
a signficantly larger value for \la\ and a stronger field dependence
(about a factor 2) than those obtained by a Lorentzian cutoff\cite{Sonier:00}.

Besides the London model, there are a couple of models for $B({\bf r})$
based on the Ginsburg-Landau theory.  Although the GL equations can be solved to
yield an approximate analytical solution for the mixed state near $H_{c1}$ or $H_{c2}$, it
must be solved numerically for intermediate fields.
Fortunately,  it is known that the field distribution obtained from 
exact numerical solutions of the GL equations is in excellent agreement with
that from the modified London model at low fields and arbitrary $\kappa$
($=\lambda/\xi_{\rm GL}$, the GL parameter)\cite{Fesenko:93}.

\begin{figure}[t]
\begin{center}
\includegraphics[width=0.65\linewidth]{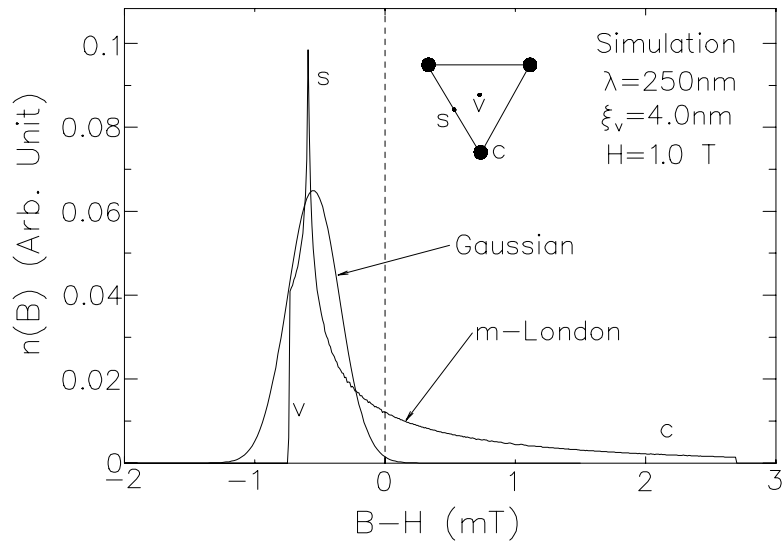}
\caption{\label{bdis} A magnetic field distribution $n(B)$ calculated by the
modified London model with a triangular flux line lattice. (The Lorentzian cutoff
was assumed.) The marking `c' refers to the contribution
from vortex cores, `s' to saddles, and `v' to valleys, respectively. The Gaussian
distribution is only meant for a guide to eye.}
\end{center}
\end{figure}

One of the most important characteristics of $n(B)$ obtained from these models
is the site-selective feature of the line shape.
As shown in Fig.~\ref{bdis}, the sharp peak due to the van Hove singularity
found in the lower field mainly represents the contribution from the saddle
points of $B({\bf r})$, the lower field end from the central valleys among
vortices, and the high field end from the region near the vortex cores.  This
quite asymmetric field profile with such a geometrical correspondence
allows us to determine $\lambda$ and $\xi_{\rm v}$ reliably by comparing
the time evolution of the muon spin polarization with that calculated by $n(B)$.

It is often the case that the superconducting properties in non-cubic
compounds are strongly anisotropic, leading to a large
difference between the magnetic penetration depths for in-plane and perpendicular
directions.
More specifically, in uniaxial superconductors with $M_{\rm ab}$ and
$M_{\rm c}$ being the carrier mass for in-plane and perpendicular directions, we have
\begin{equation}
B({\bf r},\theta)= B_0\sum_{\bf K}
b({\bf K})e^{-i{\bf K}\cdot{\bf r}}F(K,\xi_{\rm v}),
\end{equation}
\begin{equation}
b({\bf K})= \frac{1+K^2m_{zz}\lambda^2}
{(1+K_x^2m_{\rm ab}\lambda^2+K_y^2m_{xx}\lambda^2)(1+K^2m_{zz}\lambda^2)
-K^2K_y^2m^2_{xz}\lambda^4}\:,
\end{equation}
where $\theta$ is the polar angle of the $c$-axis,
$m_{\rm ab}=M_{\rm ab}/\overline{M}$,
$m_{\rm c}=M_{\rm c}/\overline{M}$
(with $\overline{M}=(M_{\rm ab}^2M_{\rm c})^{1/3}$), and
\begin{eqnarray}
m_{xx}&=&m_{\rm ab}\cos^2\theta+m_{\rm c}\sin^2\theta,\\
m_{zz}&=&m_{\rm ab}\sin^2\theta+m_{\rm c}\cos^2\theta,\\
m_{xz}&=&(m_{\rm ab}-m_{\rm c})\sin\theta\cos\theta.
\end{eqnarray}
Thus, the line shape
depends on $\lambda_{\rm ab}=\lambda\sqrt{m_{\rm ab}}$
and $\lambda_{\rm c}=\lambda\sqrt{m_{\rm c}}$ in a complex manner\cite{Greer}.

\subsection{Gaussian field distribution}

When the quality of \msr\ data is good enough to be analyzed by the above model,
we can obtain $\lambda$ and $\xi_{\rm v}$ simultaneously by directly comparing the
\msr\ time spectrum with that calculated from $B({\bf r})$.
Unfortunately, our experience shows that this is not
always the case when the sample is not a single crystal, or $\lambda$
happens to be very large, etc. so that
the characteristic features of $n(B)$ and the associated time spectra important
for such analysis are smeared out.
In such a situation, the Gaussian field distribution has been used as a convenient
analytical model, where the depolarization rate
is presumed to be given by the
second moment of the field distribution ($\lambda\gg\xi_{\rm v}$),
\begin{equation}
\langle\Delta B^2\rangle=\langle(B({\bf r})-H)^2\rangle_{\bf r},
\end{equation}
which is reflected as $T_2$ relaxation in the \msr\ line shape.
The Gaussian distribution of local fields naturally leads to
a Gaussian depolarization function,
\begin{eqnarray}
\hat{P}(t) &\simeq&\exp(-\sigma_{\rm b}^2t^2)
\exp(-\sigma^2t^2/2)\exp(i\gamma_\mu Ht+\phi),\label{gauss}\\
\sigma&=&\gamma_\mu\sqrt{\langle\Delta B^2\rangle }.
\end{eqnarray}
For the ideal case of a triangular FLL with an isotropic effective carrier mass
and a cutoff $K\approx1.4/\xi_{\rm v}$
provided by the numerical solution of the GL theory, $\lambda$
can be deduced from $\sigma$ using the following
relation\cite{Pincus:64,Aeppli:87,Brandt:88b}:
\begin{equation}
\sigma(h)\ [\mu{\rm s^{-1}}] = 4.83\times 10^4(1-h) \lambda^{-2}\ [{\rm nm}],
\label{sgmhl}
\end{equation}
where $h=H/H_{c2}$.
While the above form is valid for $h<0.25$ or $h>0.7$,
a more useful approximation valid for an arbitrary field is \cite{Brandt:88b}
\begin{equation}
\sigma(h)\ [\mu{\rm s^{-1}}] = 4.83\times 10^4
(1-h)[1+3.9(1-h)^2]^{1/2} \lambda^{-2}\ [{\rm nm}].
\label{sgmh}
\end{equation}
The field dependence of those equations represent a
reduction of the Gaussian width due to a stronger overlap of vortices at higher
fields, while $\lambda$ is a constant; therefore, the deviation of $\sigma(h)$
from those equations can be attributed to the change of $\lambda$ with the field.

However, the microscopic situation of the FLL state is considerably
different from the above ``ideal" one in practical cases where Gaussian
damping is actually observed, because there must be an additional
effect of randomness to round up the sharp feature of $n(B)$.
This also makes it difficult to distinguish
$\sigma_{\rm b}$ from $\sigma$ in Eq.~(\ref{gauss}),
giving rise to a problem in comparing the values of $\lambda$, for example,
between those from an analysis using the modified London model and those
obtained from the Gaussian approximation (where the influence of $\sigma_{\rm b}$
is indistinguishable).

%%
%In both cases, $\lambda$ is related to the superconducting carrier density
%$n_s$ as
%\begin{equation}
%\lambda^2=\frac{m^*c^2}{4\pi n_se^2},\label{lmd-ns}
%\end{equation}
%indicating that $\lambda$ is enhanced upon the reduction of $n_s$
%due to the quasiparticle excitations.

\begin{figure}[t]
\begin{center}
\includegraphics[width=0.65\linewidth]{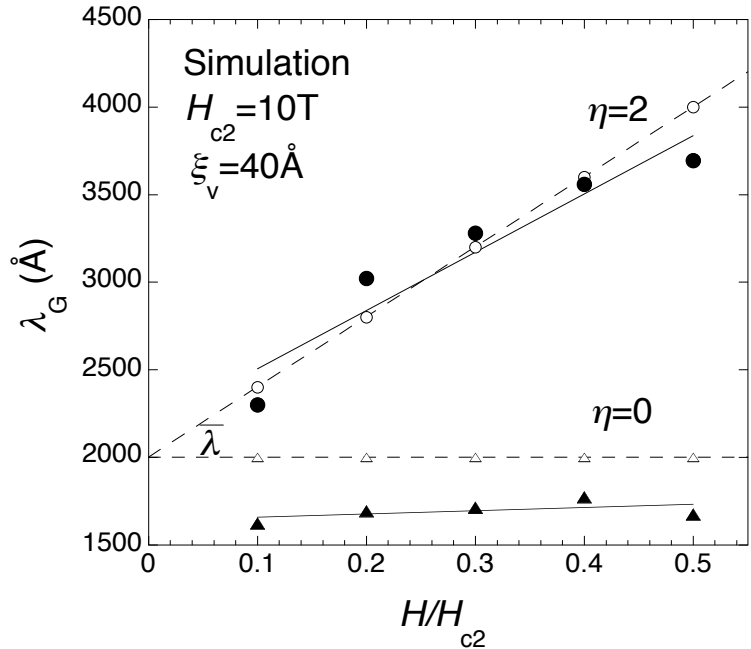}
\caption{\label{simu} Magnetic penetration depth ($\lambda_{\rm G}$)
deduced from an analysis using the Gaussian damping
(filled triangles and circles), where the analyzed
spectra were those simulated by the modified London model with  $\overline{\lambda}$
shown by the open triangles  (corresponding to $\eta=0$) and circles
 ($\eta=2$) at each field.  A linear fit yields $\eta=0.11$ and 1.53, respectively.
The upper-critical field was $H_{c2}=10$ T, from which the cutoff parameter
(\xv) was determined to be 40 \AA\ (
$=0.7\xi_{\rm GL}$, where $\xi_{\rm GL}=\sqrt{\Phi_0/2\pi H_{c2}}$,
with $\Phi_0$ being the quantum flux).}
\end{center}
\end{figure}

In order to examine the model dependence of the analysis, we made a simulation
to compare the results of an analysis, where we generated
$\mu$SR time spectra using a modified-London model, and then analyzed them
 by simple Gaussian damping (Eq.~(\ref{gauss})) to deduce $\sigma$.
According to our result, the Gaussian distribution originates
from the distribution of $\lambda$ (which may vary at different sample domains),
which must be present in the polycrystalline powder specimen of anisotropic
superconductors (e.g., $\lambda_{\rm c}\gg\lambda_{\rm ab}$).
As is also clear in the field profile shown in Fig.~\ref{bdis},  the simulated
time spectra with typical values for $\lambda$ (2000--3000 \AA)
cannot be fitted by Eq.~(\ref{gauss}) due to the strongly
exponential-like damping; this is obviously due to the
contribution of high-frequency tails in the spectral
density, $n(B)$. (Thus, the use of the second moment as an
approximation in the ideal situation would be valid only when the relaxation rate is small
enough to eliminate the asymmetric feature of $n(B)$.)
The situation was much improved when the Gaussian
distribution of $\lambda$ was introduced with a variance ($\sigma_\lambda$),
\begin{equation}
G(\lambda)\propto\exp[-(\lambda-\overline{\lambda})^2/\sigma_\lambda^2],
\end{equation}
where $\overline{\lambda}$ is the mean value.
For the parameter values shown in Fig.~\ref{simu}, the time spectra become
Gaussian-like when $\sigma_\lambda\sim$600--800 \AA, yielding reasonable
reduced chi-square values by Eq.~(\ref{gauss}). We also assumed a gradual
decrease of $\sigma_\lambda$,
\begin{equation}
\sigma_\lambda(h)=\sigma_\lambda(1-h^2),\label{sgmlmd}
\end{equation}
considering that the elastic moduli $C_{ii}$ of FLL, which control the
FLL distortion and the associated modulation of $\lambda$, exhibit a
quadratic dependence on the applied field.  (Note, however, that the
factor $1-h^2$ yields only a small change of $\sigma_\lambda(h)$ for $h<0.5$.)
Some examples are shown in Fig.~\ref{simu} for
 $\sigma_\lambda=800$ \AA, where the penetration depth ($\lambda_{\rm G}$),
obtained by Eq.~(\ref{sgmh})
from $\sigma(h)$, is plotted together with the original $\overline{\lambda}$.
A reasonable agreement between $\lambda_{\rm G}$ and
$\overline{\lambda}$ is seen, except for the case when
$\overline{\lambda}=2000$ \AA, where $\lambda_{\rm G}$ takes systematically
lower values at all fields.  This can be readily understood by considering the
fact that $\sigma$ is enhanced by an amount $\sigma_{\rm p}$ as a
consequence of Eq.~(\ref{sgmlmd}).
We also examined the
field dependence of $\lambda$, which would be most crucial in the following arguments,
\begin{equation}
\lambda(h)=\lambda(0)[1+\eta\cdot h],\label{lmdeh}
\end{equation}
where $\eta$ is a dimensionless parameter used to express the magnitude of the
field dependence.  As is evident in Fig.~\ref{simu}, the slope $d\lambda_{\rm G}/dh$
is slightly weaker than the original assumption;
when we take $\eta=2$ for the simulation,
we obtain $\eta=1.53$ as the corresponding slope for $\lambda_{\rm G}$.
However, when there is no field dependence of $\overline{\lambda}$ ($\eta=0$),
$\lambda_{\rm G}$ exhibits the least dependence on the field ($\eta=0.11$).
Thus, we can conclude that the field dependence of the penetration depth
(as a mean value) obtained from the Gaussian field approximation provides a
sound basis for the characterization of superconductors.

\section{The Doppler shift and associated non-linear effect}

\begin{figure}[t]
\begin{center}
\includegraphics[width=0.85\linewidth]{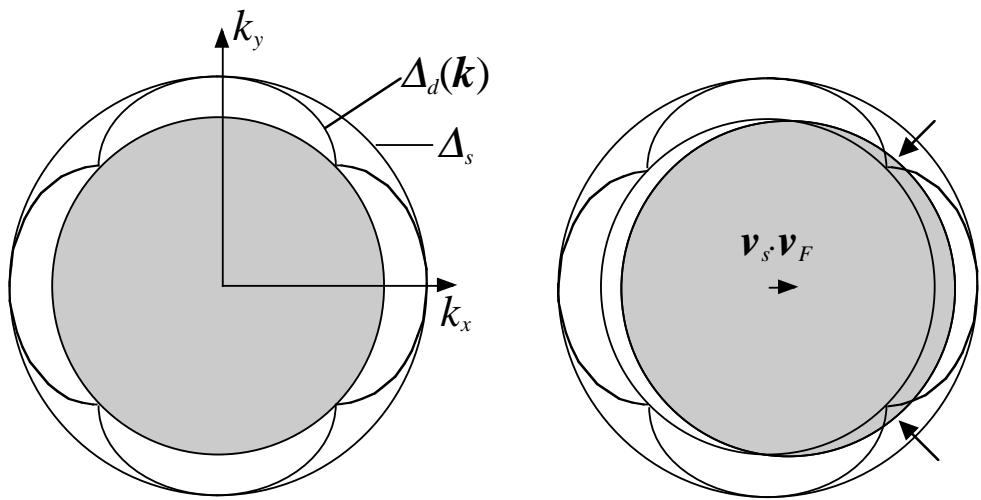}
\caption{\label{doppler} Fermi surface shifted by the quasiclassical Doppler effect 
due to the supercurrent ${\bf v}_s$ around vortices.  While such a shift has no effect on
the quasiparticle excitation for  the isotropic gap ($\varepsilon({\bf v}_{\rm F}\cdot {\bf v}_{\rm s})<\Delta_s$), it induces additional excitation by breaking pairs near to the nodal region (pointed by arrows).}
\end{center}
\end{figure}

In the FLL state, the quasiparticle momentum ${\bf v}_{\rm F}$ is shifted by the flow of
supercurrent ${\bf v}_{\rm s}$ around  the vortices due to a semi-classical Doppler shift,
leading to a shift of the quasiparticle energy spectrum to an amount $\varepsilon=m{\bf v}_{\rm F}\cdot {\bf v}_{\rm s}$.
 Since the density of state (DOS), $N(E)$, is non-zero, except at the Fermi level
($E=0$) and is higher
at larger energy ($0<E<\Delta_0$) for the anisotropic order parameter, 
quasiparticles can be excited by the Doppler shift outside of the vortex cores
with a population proportional
to $N(E+\varepsilon({\bf v}_{\rm F}\cdot {\bf v}_{\rm s}))$, leading to
an enhancement of $\lambda$ \cite{Volovik:93}. In other words, the Cooper
pairs with a gap energy of less than $\varepsilon$
can be broken by the Doppler shift (see Fig.~\ref{doppler}).
Historically, a similar effect was
considered first for type I superconductors, where the non-linear response
of the shielding current in the Meissner effect due to the ``backflow" of
quasiparticles was discussed\cite{Bardeen:54}.

\begin{figure}[t]
\begin{center}
\includegraphics[width=0.65\linewidth]{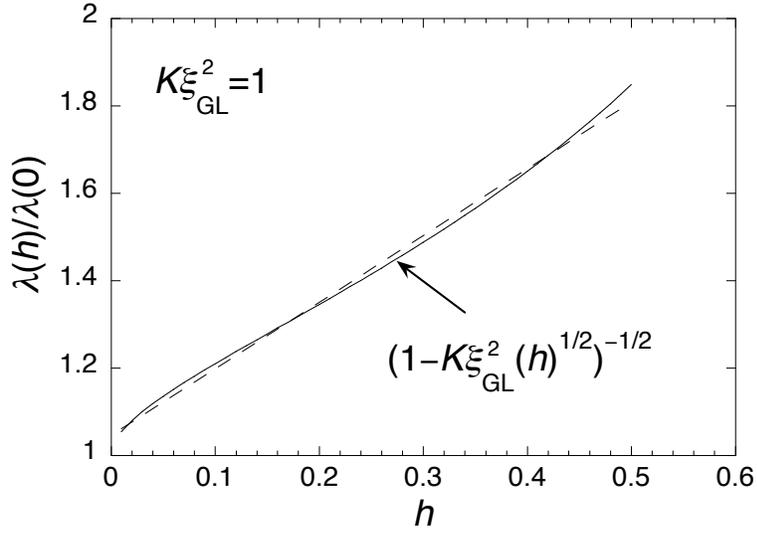}
\caption{\label{lmdnrm} Field dependence of $\lambda$ normalized
by the value at $h=0$ for the case $g(h)=\sqrt{h}$ ($K\xi_{\rm GL}^2=1$).
It is well represented by a linear dependence, as in Eq.~(\ref{lmdh}) with
$c\sim 1.5$.}
\end{center}
\end{figure}

The magnitude of $\eta$ represents the degree for the increase of DOS
for quasiparticles, which must be
roughly proportional to the phase volume of the Fermi surface
where the Doppler shift exceeds the gap energy
($\varepsilon({\bf v}_{\rm F}\cdot {\bf v}_{\rm s})>\Delta({\bf k})$).
It also follows that the effect depends on the
direction of ${\bf v}_{\rm s}$ (and hence that of the
external field ${\bf H}$ relative to the order parameter) in a sigle
crystalline specimen.
According to Volovik,
the quasiparticle density of state for the anisotropic order parameter is
\begin{equation}
N_{\rm deloc}(0)\simeq N_{\rm F} K \xi_{\rm GL}^2\sqrt{h}
\equiv N_{\rm F}g(h),\label{gh}
\end{equation}
\begin{equation}
K\propto\int_{|\Delta({\bf k})|<\varepsilon}|\Delta({\bf k})|d{\bf k},
\end{equation}
where $N_{\rm F}$ is the DOS for the normal state
and $K$ is a constant on the order of unity\cite{Volovik:93}.
It is important to note that $K$ is proportional to the phase volume
of the low excitation energy in $\Delta({\bf k})$, thereby carrying information
on the degree of anisotropy for $\Delta({\bf k})$; the factor $h^{1/2}$
comes from the inter-vortex distance ($\propto h^{-1/2}$) multiplied by 
the number of vortices ($\propto h$).
The superfluid density at a given field is then
\begin{equation}
n_{\rm s}(h)\simeq n_{\rm s}(0)[1-g(h)],
\end{equation}
which is directly reflected in the magnetic penetration depth,
\begin{equation}
\frac{1}{\lambda^2(h)}=\frac{4\pi e^2}{m^*c^2}n_{\rm s}(h).
\end{equation}
Therefore, as a mean approximation, we have
\begin{equation}
\lambda(h)=\frac{\lambda(0)}{\sqrt{1-g(h)}}
\sim\lambda(0)[1+cK\xi_{\rm GL}^2h],
\label{lmdh}
\end{equation}
where $c\simeq1.5$ for $0<h<0.5$, as show in in Fig.~\ref{lmdnrm}.
Thus, the comparison between Eqs.~(\ref{lmdeh}) and (\ref{lmdh}) yields
\begin{equation}
\eta\simeq cK\xi_{\rm GL}^2,
\end{equation}
indicating that the slope $\eta$ reflects the phase volume of
the Fermi surface where $|\Delta({\bf k})|<\varepsilon$.

Since the Doppler shift is far smaller than the gap energy in the usual situation for
the isotropic gap, no such enhancement is expected for the conventional
$s$-wave pairing ($\eta\ll$1).  A recent theoretical calculation based on
the Bogoliubov-de-Gennes (BdG) equations
indicates, however, $\eta$ is not exactly zero for $s$-wave pairing,
although it is much smaller than that for $d_{x^2-y^2}$-pairing\cite{Wang:98}.
It must also be noted that the effect of temperature must be considered
to evaluate the degree of anisotropy from the measurement of $\eta$.
In general, the phase volume of the Fermi surface where $|\Delta({\bf k})|<k_{\rm B}T$
also contributes to quasiparticle excitation. Thus, strictly speaking,
the observation of a finite $\eta$ means the presence of a small gap
region with an upper bound of $\sim k_{\rm B}T$ (see below, the case of
MgB$_2$ for example).

\section{Non-local corrections}

Because a superconducting pair correlation occurs over a finite length scale,
$\xi_0$ (i.e, the BCS coherence length), the electromagnetic response of
superconductors is subject to various non-local effects.  The primary example
is the cutoff term, $F(K,\xi_{\rm v})$, incorporated in the modified London
model (Eq.~(\ref{Br})).  Moreover, there are a couple of other corrections
that must be considered for anisotropic superconductors.

In the superconducting state with gap nodes in the order parameter,
the quasiparticles are mostly confined to the vicinity of the nodes at low
temperatures. This generally leads to the suppression of quasiparticle
excitation due to the non-local electrodynamics caused by the divergent
coherence length, $\xi\propto\Delta({\bf k})^{-1}$,
yielding a weaker temperature/field dependence of
$\lambda$ at higher fields\cite{Kosztin:97,Amin:98}.  Such an effect
has been studied experimentally in detail for the case of
YBa$_2$Cu$_3$O$_{6.95}$\cite{Sonier:99,Sonier:00}.

Another important correction comes from the anisotropy of the Fermi surface.
Novel superconductors including cuprates have a common feature that
the Fermi surface tends to exhibit a strong anisotropy due to, e.g.,  a two-dimensional
and/or multi-band character, which influences the flow of supercurrent over a length scale,
$\hbar{\bf v}_{\rm F}/\Delta_0$\cite{Affleck:97,Kogan:97a,Kogan:97b,Agterberg:98}. More specifically, the length scale is
also controlled by the mean-free path ($l$) for electrons.  It turns out that
this non-local correction partially accounts for the change in the vortex lattice structure
from triangular to squared lattice in various systems, including
$R$Ni$_2$B$_2$C
($R$=Y, Lu)\cite{Paul:98,Yethiraj:98,Gammel:99,Sakata:00,Eskildsen:01,Ohishi:02},
V$_3$Si\cite{Yethiraj:99, Sosolik:03}, and
La$_{2-x}$Sr$_x$CuO$_4$ ($x=0.17$)\cite{Gilardi:02}.
This also leads to a change in $B({\bf r})$ due to the modified flow of
supercurrent from circular to squared shape (see below, e.g. YNi$_2$B$_2$C).
Note, however, that the difference
in the free energy between a triangular and a square vortex lattice is fairly small,
making the lowest-energy configuration strongly dependent on other physical
quantities, such as the temperature, magnetic field, and crystal orientation, which is
also in a strong correlation with the superconducting order parameter.

\section{Overview of $\mu$SR results}

In this section, we try to establish the correspondence between the presence of
the anisotropic order parameter and a non-zero slope ($\eta$) in the magnetic-field 
dependence of $\lambda(h)$.  As shown below, $\eta$ provides a
good measure for the degree of anisotropy in the superconducting order
parameter.  However, one has to keep in mind that additional
information is generally needed to resolve the precise symmetry of the
pairing; one would be easily led to a false conclusion in choosing, e.g., between
$d$- and $s+g$-wave pairing based solely upon the $\mu$SR result.
Another origin of apparent anisotropy would be multi-gap
order parameters with one of those having a small gap ($<k_{\rm B}T$),
as suggested in the case of MgB$_2$ (see below).

While a vast body of superconductors
have been investigated by $\mu$SR, there are not many of them in which
the field dependence of $\lambda(h)$ has been measured in detail.  This is partly
due to the historical reason that the $\mu$SR apparatus with a high magnetic
field has become available for routine service only since the late '90s.
Here, we summarize our recent work on CeRu$_2$,
Y(Pt,Ni)$_2$B$_2$C, Cd$_2$Re$_2$O$_7$,
and MgB$_2$, in which detailed $\mu$SR measurements have been performed.
The results of NbSe$_2$ and YBa$_2$Cu$_3$O$_{6.95}$ are also mentioned
for a comparison.  A wider variety of compounds in view of other experimental
techniques are covered elsewhere\cite{Brandow:03}.

\subsection{CeRu$_2$}
The cubic Laves phase compound,  CeRu$_2$ ($T_c=6.1$--6.5 K at zero
field, $H_{c2}(T=0)\simeq 6$--7 T), has a long history of experimental
and theoretical studies since its discovery in 1950s.
One of the current issues is its magnetic response at higher fields ($h>0.5$),
where an anomalous enhancement of quasiparticle excitation has been reported.
This is suggested by the
observation of de Haas-van Alphen (dHvA) oscillation over a field region where
the cyclotron radius is much larger than the inter-vortex distance\cite{Hedo:98a}.
The presence of excess quasipaticles has been further 
confirmed by a strong enhancement of \la\ measuremed by \msr\ at
higher fields\cite{Yamashita:97}.
Moreover, while most of experimental studies concluded that
the pairing symmetry is a spin-singlet $s$-wave, detailed studies on the spin-lattice
relaxation in
nuclear quadrupole resonance (NQR) suggested the presence of anisotropy
in the order parameter\cite{Mukuda:98}.  This was apparently in line
with the observed non-linear field dependence ($\propto h^{1/2}$)
of the electronic specific heat
coefficient $\gamma(h)$\cite{Hedo:98b};
As indicated in Eq.~(\ref{gh}), the quasiparticle excitation has a contribution
proportional to $h^{1/2}$ for the anisotropic order parameter, while
an $h$-linear dependence is
expected for the conventional case, because $\gamma(h)$ must be
proportional to the volume of normal cores, and thereby to the number of vortices.

\begin{figure}[t]
\begin{center}
\includegraphics[width=0.65\linewidth]{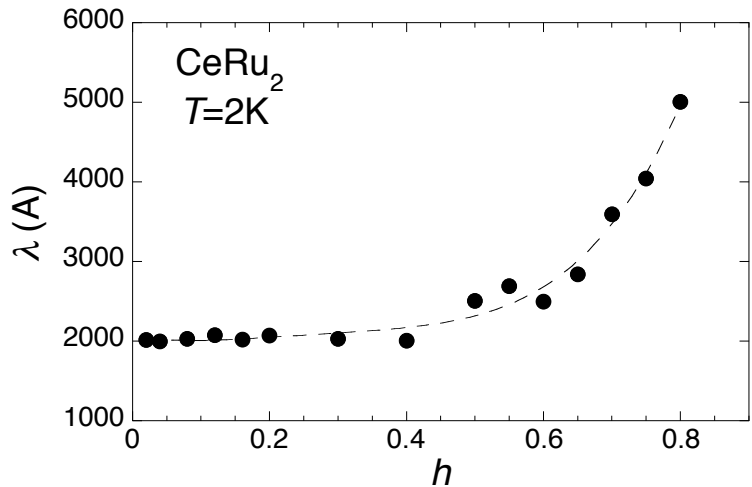}
\caption{\label{lmdceru} Field dependence of the magnetic penetration depth
($\lambda$) in CeRu$_2$ at 2 K obtained by fitting data with the modified
London model, where the dashed line is a guide for the eye 
(after Ref.\cite{Kadono:01}).}
\end{center}
\end{figure}

However, our $\mu$SR studies in a high quality single crystal has shown that
the field dependence of $\gamma(h)$ can be attributed to that of the
vortex core radius $\rho_{\rm v}$ ($\propto\xi_{\rm v}$),
at least over the region $h<0.5$;
the $\mu$SR spectra were analyzed (in the frequency domain) by the
modified London model to extract $\xi_{\rm v}$
and $\lambda$ independently\cite{Kadono:01}.  (We later re-analyzed a part
of the data in time domain and found that the result was unchanged.)
When the quasiparticles are confined within the vortex cores, the
electronic specific-heat coefficient must be proportional to the cross
section of the cores multiplied by the number of vortices, 
\begin{eqnarray}
\gamma(h)\propto \pi \rho_{\rm v}^2(h)\cdot h= \pi \rho_0^2 h^\beta,\\
\rho_{\rm v}(h)=\rho_0 h^{(\beta-1)/2},
\end{eqnarray}
where our result for $\rho_{\rm v}(h)$ is well reproduced by $\beta\simeq0.53$.
Thus, the observed nearly $h^{1/2}$ dependence of $\gamma(h)$\cite{Hedo:98b} 
exhibits perfect agreement with our result. This is also qualitatively
in line with a recent theoretical calculation for $s$-wave superconductors based on the
quasi-classical Eilenberger equations, where $\beta\simeq0.67$ is predicted
due to various non-trivial effects, including
those from the inter-vortex interaction and the vortex core excitation at
finite temperature\cite{Ichioka:99}.
This strongly suggests that quasiparticle excitation is confined
within the normal cores of the vortices.
More importantly, we found that $\lambda(h)$ exhibits the least dependence on $h$
(namely, $\eta\simeq0$)  over the relevant field range, as clearly shown in 
Fig.~\ref{lmdceru}\cite{Kadono:01}.
This is perfectly in line with the above conclusion obtained for the
vortex cores, as is also the case with $s$-wave pairing suggested by the vast
majority of other experimental results.

Meanwhile, it must be stressed that the situation changes drastically upon
increasing the field above $h\sim0.5$.  Figure \ref{lmdceru} indicates that
$\lambda$ exhibits a divergent increase for $h>0.6$ with increasing field, thereby suggesting
a divergent increase of quasiparticle excitation.  This is strongly
supported by the observation of the dHvA effect, as mentioned above.  To our
knowledge, there is no simple explanation of such an anomaly\cite{comm}.
One possible model
may be that proposed by Fulde-Ferrel-Larkin-Ovchinnikov, where a new superconducting
phase with a spatially inhomogeneous order parameter is
predicted to occur in rare-earth compounds having a large spin
paramagnetism\cite{Fulde:64,Larkin:65}.

Finally, we note that the absence of a clear coherence peak in NQR may be
due to the weak random magnetism observed  below $\sim$40 K 
by zero field $\mu$SR\cite{Huxley:96}. They observed an increase in the muon 
spin relaxation rate on the order of 0.02$\mu$s$^{-1}$ in accordance with the 
increase of the $ac$-susceptibility\cite{Nakama:95}.  
The sample quality suggests that the weak magnetism is of intrinsic origin, which 
would act as a scattering source for pair breaking.

\subsection{Y(Ni,Pt)$_2$B$_2$C}

The borocarbide superconductor, YNi$_2$B$_2$C
($T_c=15.4$ K at zero
field, $H_{c2}(T=0)\simeq 7$--8 T),  has attracted much attention
due to the strong $h^{1/2}$ dependence of the electronic specific-heat
coefficient in a high-purity specimen\cite{Nohara:99},
with which they suggested a change of the vortex core radius based on
earlier experimental suggestions that an $s$-wave pairing was
realized in this compound; the system showed little sensitivity to
non-magnetic impurities, as is typically found in BCS  $s$-wave
superconductors.  They also found
that such a $h^{1/2}$ dependence was replaced by a $h$-linear dependence
upon substitution of Ni by Pt ($\simeq 20$\%).
By now, there is mounting evidence that the order parameter in
pure YNi$_2$B$_2$C is considerably
anisotropic\cite{Terashima:97,Yang:00,Boaknin:01,Izawa:01,Izawa:02,Lipp:02,Park:03},
although the pairing symmetry
is basically $s$-wave-like.  The key to understand the varying results
in borocarbides is that the anisotropy of the order parameter is
indeed sensitive to impurities, which is in good contrast to the robustness
of superconductivity, itself; $s+g$-pairing, for example, changes
into an effectively isotropic $s$-wave pairing, where the anisotropic part 
(associated with $g$-component) is washed 
out by impurity scattering.
Thus, a part of the divided results obtained by NMR-$1/T_1$
measurements may be sorted out in terms of the sample  purity\cite{Brandow:03}.
Our $\mu$SR study was quite successful to clarify the effect of non-magnetic 
impurities on the anisotropic order parameter in Y(Ni$_{1-x}$Pt$_x$)$_2$B$_2$C.

As mentioned before, there is a significant contribution of the non-local
effect in borocarbides due to the anisotropy of the Fermi surface.  This effect
must be considered for modeling of the magnetic field distribution,
$B({\bf r})$, and the associated spectral density, $n(B)$, to obtain the 
correct values for $\lambda$ and $\xi_{\rm v}$.  To this end, the London model
is further modified to yield
\begin{equation}
B({\bf r})=B_0\sum_{\bf K}\frac{e^{-i{\bf K}\cdot{\bf r}}}
{1+K^2\lambda^2+(c_1K^4+c_2K_x^2K_y^2)\lambda^4}F(K,\xi_{\rm v})\:,
\label{Brnl}
\end{equation}
where the terms proportional to $K_x^2K_y^2$ represent the non-local
effect, with $c_i$ being the parameters coming from the band structure\cite{Kogan:97a}. 
Moreover, the non-local effect leads to the formation of a squared flux line lattice,
which also modifies $B({\bf r})$.   Our result indicates that
these features have a strong influence on $n(B)$ probed by $\mu$SR\cite{Ohishi:02}.
For example, no reasonable fit can be obtained when one assumes a square FLL 
without non-local correction terms.

\begin{figure}[t]
\begin{center}
\includegraphics[width=0.65\linewidth]{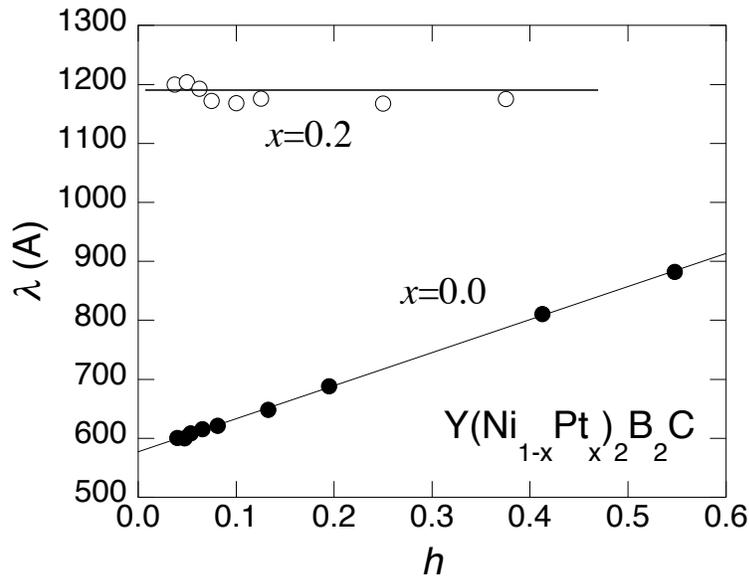}
\caption{\label{lmdynbc} Field dependence of the magnetic penetration depth
($\lambda$) in Y(Ni$_{1-x}$Pt$_x$)$_2$B$_2$C, obtained by fitting data with 
the modified London model and a non-local correction, where the data were
obtained at 3 K for $x=0.0$ and 2.5 K for $x=0.2$ ($T_{\rm c}=12.1$ K), respectively
(after Ref.\cite{Ohishi:02,Ohishi:03a}).}
\end{center}
\end{figure}

Our $\mu$SR experiment has revealed that the vortex core radius, $\rho_{\rm v}$,
in a pure specimen ($x=0$) exhibits a much
steeper decrease with increasing field than that estimated from 
the electronic specific-heat coefficient, $\gamma(h)$; $\rho_{\rm v}(h)$ shrinks sharply
for $h<0.1$, then changes only very weakly with the field.  This is in marked contrast
with the case of CeRu$_2$, where the field dependence of $\gamma(h)$ is
in good accord with $\rho_{\rm v}(h)$ 
(i.e., $\gamma(h)\propto h\pi\rho_{\rm v}^2(h)$).
This suggests the presence of excess quasiparticle excitation outside of the vortex cores,
which contributes to $\gamma$.
As shown in Fig.~\ref{lmdynbc}, this is indeed supported by the observation that 
$\lambda$ exhibits a strong field dependence in a pure specimen.
The slope is deduced from the linear fitting (Eq.~(\ref{lmdeh}))
to yield $\eta=0.95(1)$.  On the other hand in the Pt-doped specimen ($x=0.2$), 
$\lambda$ is mostly independent of the field ($\eta\simeq0$)\cite{Ohishi:03a}.  
This is again
consistent with the presumed hybrid nature of the order parameter, where
the $s$-wave component is relatively enhanced by impurity scattering.
We also note that $\lambda$ in the Pt-doped specimen is about 1.27-times
longer than that expected solely by the conventional impurity effect\cite{Ohishi:03a}.
This strongly suggests that there is an excess quasiparticle density of states
generated by the interaction between the impurities and the anisotropic component
of the order parameter, as is found in superconductors with gap nodes.

\subsection{MgB$_2$}
The revelation of superconductivity in a binary intermetallic compound,
MgB$_2$, has attracted much interest because it exhibits an almost two-times
higher transition temperature  ($T_{\rm c}\simeq39$ K) than those of all
intermetallic superconductors known to date\cite{Nagamatsu:01}.
The most interesting issue concerning this compound is whether or not it
belongs to the class of the conventional BCS type (namely, phonon-mediated
spin-singlet $s$-wave pairing) superconductors.
So far, most experimental results favor phonon-mediated 
superconductivity\cite{Budko:01,Takahashi:01,Kotegawa:01,XKChen:01,Karapetrov:01,Sharoni:01,Schmidt:01,Pronin:01}.
On the other hand, calculations of the band structure and the phonon spectrum 
predict a double energy gap \cite{Kortus:01,Liu:01}, with a larger gap attributed 
to two-dimensional $p_{x-y}$ orbitals, and a smaller gap to 
three-dimensional $p_z$ bonding and antibonding orbitals. 
The experimental results of specific heat measurements \cite{Bouquet:01,Wang:01}, 
point-contact spectroscopy \cite{Szabo:01}, photoemission spectroscopy 
\cite{Tsuda:01}, 
scanning tunneling spectroscopy \cite{Giubileo:01} and penetration depth 
measurements \cite{Manzano:01} have supported this scenario.

\begin{figure}[t]
\begin{center}
\includegraphics[width=0.65\linewidth]{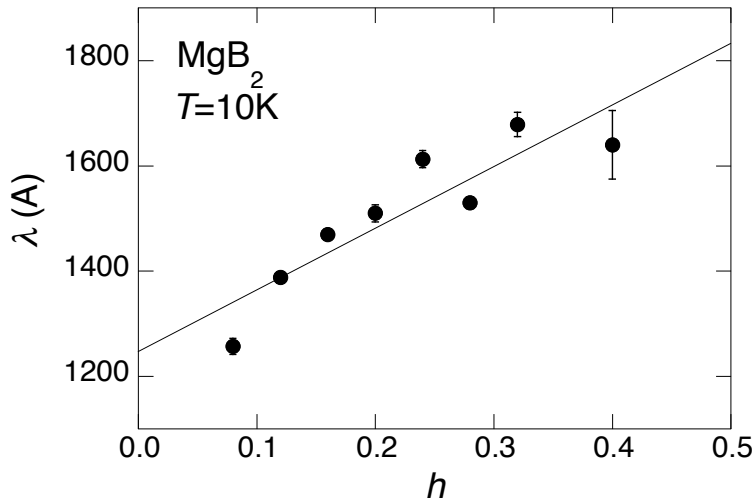}
\caption{\label{lmdmgb} Field dependence of the magnetic penetration depth
($\lambda$) in MgB$_2$ obtained by fitting data with 
the Gaussian field distribution model, where the data were
obtained at 10 K.  The values for $h<0.1$ are probably underestimated 
due to addtional relaxation induced by flux pinning
(after Ref.\cite{Ohishi:03b}).}
\end{center}
\end{figure}

The double energy gap would have a direct relevance on the temperature dependence
of $\lambda$, because there must be excess quasiparticles excited over the
smaller energy gap ($\Delta_{\rm S}$) at higher temperatures, while the bulk
superconductivity is maintained by the larger gap ($\Delta_{\rm L}$).  
At this stage, there are two such
$\mu$SR measurements reporting the result of an analysis based on the two-gap
model, where $\Delta_{\rm S}$ is reported to be 2.6(2) meV\cite{Niedermayer:02}
and 1.2(3)  meV\cite{Ohishi:03b}, respectively.
On the other hand, the field dependence
of $\lambda$ is sensitive only to those excited by the Doppler shift, and therefore
the slope $\eta$ would be zero as long as both energy gaps are isotropic.
The only exception would be that the temperature at which $\lambda(h)$
is measured is comparable to $\Delta_{\rm S}/k_{\rm B}$, so that the 
smaller gap effectively  becomes equivalent with nodes in the order parameter.
As shown in Fig.~\ref{lmdmgb}, our result of $\lambda(h)$ exhibits a
clear dependence on $h$ with $\eta=1.27(29)$,
where the measurements were performed at $T\simeq10$ K\cite{Ohishi:03b}.
Considering that the measured temperature nearly corresponds to 
$\Delta_{\rm S}/k_{\rm B}=14(4)$ K,  the observed field dependence
would be qualitatively consistent with the two-gap model with isotropic
order parameters. Meanwhile, if the smaller gap is as large as 2.6 meV\cite{Niedermayer:02}, it would mean that either $\Delta_{\rm L}$ or $\Delta_{\rm S}$ is
anisotropic.

Unfortunately, so far it is difficult to obtain a single crystal of 
MgB$_2$ with the dimensions necessary for the conventional $\mu$SR experiment; 
thus, all of the $\mu$SR measurements have been performed on powder specimens.
The obtained time spectra were fitted by the Gaussian field distribution 
(Eq.~(\ref{gauss})), where the additional relaxation due to the flux pinning 
($\propto\exp(-\sigma_p^2t^2$)) was not separable.
As a matter of fact, we observed an {\sl increase} of the relaxation rate 
with increasing field over the region $h<0.1$ in our MgB$_2$ specimen, which 
might be related to the flux pinning\cite{Ohishi:03b}.
However, as we showed in Section 2.2,
an analysis based on the Gaussian field distribution model has a relatively
weak uncertainty in terms of the {\sl relative} change of $\lambda$
against the field.  Thus, we think that the above result provides a sound basis
for a qualitative evaluation of the gap anisotropy.

Another source of complication would be that there is a strong anisotropy of magnetic
response over the crystal direction, which is mixed up in the polycrystalline
specimen; it has been reported that the upper critical field 
for $H$ parallel to the $ab$-plane ($H_{c2}^{\rm ab}$)
is about three-times higher than that for the perpendicular direction
($H_{c2}^{\rm c})$\cite{Shi:03}.  This introduces an uncertainty in the
definition of the normalized field $h$ ($=H/H_{c2}$), which is directly
reflected in the evaluation of $\eta$.  Thus, further measurements on a single crystalline specimen would be necessary for the reliable evaluation of $\eta$.

Recently, we made $\mu$SR measurements on 
a new superconductor, Ca(Si$_{0.5}$Al$_{0.5}$)$_2$
($T_c=7.7$ K), which has a crystal structure quite similar to that of 
MgB$_2$\cite{Imai:02}.  Provided that the structure of order parameter
in MgB$_2$ is strongly related with that of the Fermi surface, a similar situation
might be expected in this compound.
Our preliminary result on a polycrystalline
specimen with the Gaussian analysis indicates that
$\lambda$ exhibits a field dependence with $\eta\simeq1.85$, 
thereby supporting the above conjecture, at least in terms of quasiparticle
excitation\cite{Kuroiwa:04}.

\subsection{Cd$_2$Re$_2$O$_7$}

A class of metal oxides isostructural to mineral pyrochlore 
has been attracting considerable attention because they exhibit a wide variety of
interesting physical properties\cite{Subramanian:83}. 
The pyrochlore has a general formula of A$_2$B$_2$O$_7$,
consisting of BO$_6$ octahedra and
eightfold coordinated A cations, where A and B are
transition metals and/or rare-earth elements.  In particular,
the B sublattice can be viewed as a three-dimensional network
of corner-sharing tetrahedra, providing a testing ground for studying
the role of geometrical frustration in systems that have local spins at 
B sites with an antiferromagnetic (AFM) correlation.\cite{Ramirez:94}
Although metallic pyrochlores comprise a minority subgroup of the pyrochlore family,
they consist of distinct members, such as Tl$_2$Mn$_2$O$_7$,
which exhibits a colossal magnetoresistance.\cite{Shimakawa:96,Shimakawa:99}
In view of these backgrounds, the recently revealed superconductivity in 
5$d$ transition metal pyrochlores and related oxides, 
Cd$_2$Re$_2$O$_7$\cite{Hanawa:01,Jin:01}
and KOs$_2$O$_6$\cite{Yonezawa:04}, 
is intriguing, because they evoke anticipation for exotic superconductivity.

\begin{figure}[tb]
\begin{center}
\includegraphics[width=0.65\linewidth]{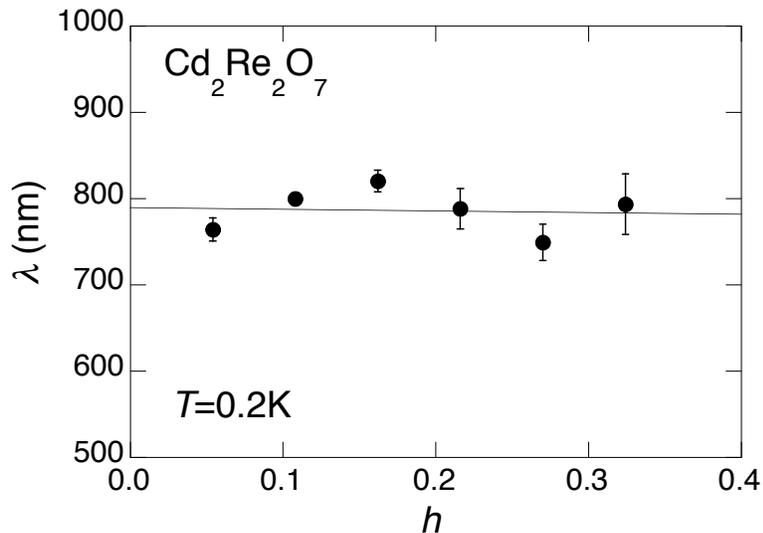}
\caption{\label{lmdcro} Field dependence of the magnetic penetration depth
($\lambda$) in Cd$_2$Re$_2$O$_7$ obtained by fitting data with 
the Gaussian field distribution model, where the data were
obtained at 0.2 K
(after Ref.\cite{Kadono:02}).}
\end{center}
\end{figure}

It is reported that Cd$_2$Re$_2$O$_7$ falls into the bulk superconducting state below 
$T_c\simeq 1\sim 2$ K\cite{Hanawa:01}.  The $dc$-magnetization curve indicates
that the superconductivity is of type II with the upper critical field close to
0.29 T at 0 K.  So far, $^{187}$Re NMR-$1/T_1$ exhibits a clear coherence
peak typically found for the conventional $s$-wave pairing\cite{Vyaselev:02},
although there are not many reports concerned with the pairing symmetry.
Thus, the existing evidence strongly suggests that the order parameter is
unexpectedly isotropic.   This is further supported by the absence of a
field dependence for $\lambda(h)$.
Figure \ref{lmdcro} shows $\lambda(h)$ versus $h$, where one can clearly
see that $\eta\simeq 0$ over the observed field range\cite{Kadono:02}.  
Here, we note
that the upper-critical field can be determined by a $\mu$SR measurement
when it is well within the reach of the apparatus ($<7$ T).  
As is evident in Eqs.~(\ref{sgmhl}) and (\ref{sgmh}), 
the spin relaxation due to FLL is quenched
at $h=1$ (i.e., $H=H_{c2}$).  Thus, from the field dependence of $\sigma$, we 
obtained  $H_{c2}=0.37(5)$ T for our specimen.  The normalized field in 
Fig.~\ref{lmdcro} is defined by this value for $H_{c2}$.  The fact that $\lambda$
exhibits the least dependence on $h$ also means that the field dependence of
$\sigma$ is well reproduced by Eq.~(\ref{sgmhl}) or (\ref{sgmh}) without
considering the change in $\lambda$ with the field.  A similar field dependence
for $\sigma$ is also reported from another group\cite{Lumsden:02}.

On the other hand, in KOs$_2$O$_6$ ($T_c\simeq9.5$ K), 
our preliminary $\mu$SR data on a powder
specimen exhibits a strong field dependence of $\lambda(h)$\cite{Koda:04},
suggesting the presence of anisotropy in the order parameter.  This is also consistent
with the absence of coherence a peak in $^{39}$K NMR-$T_1$\cite{Takigawa:04}.

\subsection{Other examples}

\begin{figure}[t]
\begin{center}
\includegraphics[width=0.8\linewidth]{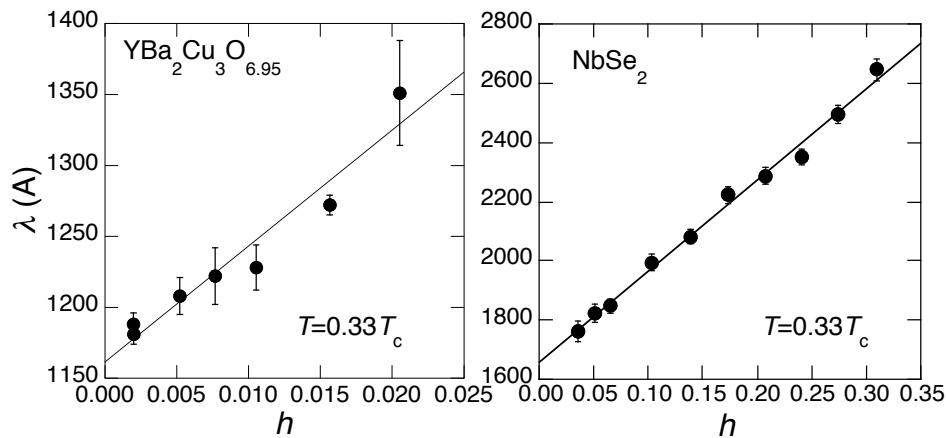}
\caption{\label{lmdybco} Field dependence of the magnetic penetration depth
($\lambda$) in YBa$_2$Cu$_3$O$_{6.95}$ (YBCO) and NbSe$_2$ at $T=0.33T_c$, 
obtained by fitting with the modified London model, where $H_{c2}=95$ T and 2.9 T, 
respectively. A linear fit yields $\eta=5$--6.6 in YBCO and 1.85(7) in NbSe$_2$.
A slightly smaller value ($\eta=1.62(6)$) is reported for the latter with an analysis
by the GL model
(after Ref.\cite{Sonier:00}).}
\end{center}
\end{figure}

As discussed in Section 3, quasiparticle excitation due to the Doppler
shift is predicted to be stronger for a larger degree of manifoldness 
in the nodal structure of the order parameter. Thus, it is naturally expected
that superconductors with $d$-wave paring  would exhibit a strong
field dependence of $\lambda(h)$.  This was proven to be the case
by systematic $\mu$SR studies on the vortex state of 
high-$T_{\rm c}$ cuprates\cite{Sonier:00}.
A typical example is found in YBa$_2$Cu$_3$O$_{6.95}$ (see Fig.~\ref{lmdybco}), 
in which $\eta$ is reported to be
5--6.6 over a field range $0<H<2$ T\cite{Sonier:97a}.  They later extended
the measurement up to 7 T, where they found that the field dependence of 
$\lambda(h)$ became weaker at higher fields 
($\eta\sim 2$ for $H>2$ T)\cite{Sonier:99}.  This is now understood
to be a consequence of the non-local correction discussed earlier in Section 4.
It must be noted that the non-local correction has a strong influence on
the temperature dependence of $\lambda$ at higher fields.  These 
results imply that one must be careful about the field range of
measurements to evaluate the meaning of $\eta$ as well as the
slope against temperature, $d\lambda/dT$.

The last example is NbSe$_2$\cite{Sonier:97b,Sonier:00}, 
where the situation is similar to that in
YNi$_2$B$_2$C or MgB$_2$.  It has been suggested from the non-linear field dependence
of $\gamma(h)$ that there must be excess quasiparticles induced by
a magnetic field\cite{Nohara:99}.  However, the degree of non-linearity 
is considerably weaker than that observed in YNi$_2$B$_2$C, suggesting
the smaller anisotropy in the order parameter.  This has been supported
by other experiments showing that NbSe$_2$ has an anisotropic
$s$-wave gap with $\Delta_{\rm max}/\Delta_{\rm min}\simeq
2$\cite{Hess:90,Sanchez:95,Ishida:96}.   
As shown in Fig.~\ref{lmdybco}, the absence of nodes, however,
does not necessarily mean $\eta=0$ when the order parameter is anisotropic (or
multi-gapped).  It happens that a temperature of $\sim0.33T_c$ (where 
measurements were performed)\cite{Nohara:99,Sonier:97b} is relatively
high, so that it is almost comparable to $\Delta_{\rm min}$.  Thus, as
discussed earlier, the region around $\Delta_{\rm min}$ in the Fermi
surface works effectively as nodes at such a high temperature.
Given that this ``quasi-node" scenario is correct, one can predict
a larger non-linearity in $\gamma(h)$ as well as a larger $\eta$
at higher temperatures.  Such a tendency is actually reported in Ref.\cite{Sonier:97b},
where they observed larger $\eta$ at $T=0.6T_c$.

\section{Summary and conclusion}

We demonstrated that the field dependence of the magnetic penetration depth
$\lambda(h)$ provides a sensitive probe for quasiparticles induced
by the Doppler shift.   As summarized in Table 1, the slope $\eta$ is positive 
when the superconducting order parameter has nodes (or a small gap
equivalent to the node at a given temperature), while it is 
close to zero for the conventional isotropic order parameter.
Despite the ambiguity associated with the slight dependence
of $\lambda$ on the employed model for data analysis, the magnitude of 
$\eta$ provides a good measure for the degree of anisotropy.
This would be useful in selecting the paring symmetry and the associated
mechanism of superconductivity for newly discovered materials.

\begin{table}[ht]
\caption{Dimensionless parameter, $\eta$, corresponding to the slope of 
$\lambda(h)$ against an external field obtained by \msr. The column `model' shows that
of the field distribution used for each analysis: `m-L' for the modified London model and
`G' for the Gaussian field distribution. $T$ denotes the
temperature where the field dependence of $\lambda$ was measured.
The values for Nb$_3$Sn are quoted from our preliminary report\cite{Kuroiwa:02}.}
\begin{tabular}{lcccccc}
\hline
\hline
 & $T_c$ (K) & pairing symmetry &  $\eta$ & model & $T$ (K) &  $H_{c2}(T)$ (T) \\
\hline
YNi$_2$B$_2$C & 15.4 & anisotropic $s$ ($s+g$?) & 0.95(1) & m-L & 3.0 & 7.0\\
MgB$_2$ & 39 & double gap & 1.3(3) & G & 10.0 & 12.5\\
NbSe$_2$ & 7.0 & anisotropic $s$ & 1.85(7) & m-L & 2.3 & 2.9\\
YBa$_2$Cu$_3$O$_{6.95}$ & 93.2 & $d$ & 5--6.6 & m-L & 31.0 & 95 \\
\hline
CeRu$_2$ & 6.0 & isotropic $s$ ($h<0.5$) & $\simeq0$ & m-L & 2.0 & 5.0\\
Y(Ni$_{0.8}$Pt$_{0.2})_2$B$_2$C & 12.1 & isotropic $s$ & $\simeq0$ & m-L & 2.5 & 4.0\\
Cd$_2$Re$_2$O$_7$ & 1--2 & isotropic $s$  & $\simeq0$ & G & 0.2 & 0.37\\
Nb$_3$Sn & 18.3 & isotropic $s$  & $\simeq0$ & G & 2.1 & $\sim$24\\
\hline
\hline
\end{tabular}
\end{table}

\ack
We thank J E Sonier for providing the numerical data in Fig.~\ref{lmdybco} and
for helpful discussions.  
The $\mu$SR experiment was
partially supported by the Grand-in-Aid for Scientific Research on Priority Areas
and the Grand-in-Aid for Creative Scientific Research from the Ministry of Education,
Culture, Sports, Science and Technology, Japan.

\end{document}